\documentstyle[twoside,fleqn,espcrc2]{article}
\begin{document}
\title{Alternative Solution of Strong CP}

\author{P.H. Frampton \address{University of North Carolina,
        Chapel Hill, NC 27599-3255}}

\begin{abstract}
In this talk I begin with some general discussion
of the history of CP violation, then move on to aspects of
a spontaneous CP violation model including the production of new particles at LHC,
implications for B decay, generalized Cabibbo mixing
and a reevaluation of kaon CP violation. Finally there is a summary.
\end{abstract}

\maketitle

\section*{Introduction}

My talk is the only non-axion talk at this Workshop but its
inclusion shows the broadmindedness of Pierre Sikivie. The only
objection I have to the axion work of Peccei and Quinn is its
non-uniqueness. I'll describe extensions of the Standard Model
which solve the strong CP problem without an axion, and have
the advantage of addressing CP violation. Due to the time available,
I will be able to give just an impressionistic view.

\section{History}

The parity operation is a symmetry of Newton's Laws provided we assume a strong
form of the Third Law: Action and Reaction are equal and opposite and directed
along the line of centers. For quantum mechanics, Parity was introduced by Wigner in 
1927\cite{wignerP}. The violation of P was first entertained by Lee and Yang
in 1956\cite{lee}; it was quickly verified by Madame Wu\cite{wu} and others
\cite{ledermantelegdi}.

Time reversal T is an invariance of Newton's Laws. In quantum mechanics T
was introduced as the now-familiar anti-unitary operator by Wigner\cite{wignerT}.
[T violation was studied in classical statistical mechanics earlier by Boltzmann
and Panlev\'{e}, but T violation in microscopic
laws was not seriously questioned until 1964.]

The operation of charge conjugation (C) could hardly be conceived of before
the Dirac equation\cite{dirac} in 1928 predicted the $e^+$, discovered in 1932.
The C invariance of quantum electrodynamics was first discussed by Kramers\cite{kramers}
in 1937. 

The invariance under CPT was proven for quantum field theory in 1954 by
Luders\cite{luders} under the weak assumptions of lorentz invariance and the spin-statistics
connection.

After Lee and Yang, but before P violation was discovered, Landau\cite{landau}
suggested that CP is an exact symmetry.

In \cite{fitch} CP violation was discovered in the decay of neutral 
kaons. The longer-lived CP eigenstate $K_L$ was observed to decay 0.2\%
of the time into $\pi\pi$, disallowed if CP is exact.
The CP violation is characterized by the parameter
$\epsilon$. Since CP violation has never been seen
outside of the kaon system, $\epsilon$ is the only
accurately measured (to within 1\%) CP violation parameter.

In a remarkable paper containing an all-time favorite idea
in particle theory, in 1966 Sakharov\cite{sakharov} proposed that the baryon number of the universe
arose due to a combination of three ingredients: (1) B violating
interactions. (2) Thermodynamic disequilibrium. (3) CP Violation.

When GUTs became popular, Yoshimura\cite{yoshimura} and others illustrated this
idea. More recently baryogenesis at the electroweak phase transition is
discussed based on the same three ingredients.

In 1973, Kobayashi and Maskawa(KM)\cite{KM} proposed their mechanism for CP
violation assuming, with great foresight, three fermion generations. The issue 
now is 
whether KM is the full explanation of the observed CP violation.

In 1976 't Hooft\cite{hooft} emphasised the strong CP problem
that a parameter $\bar{\theta}$ in QCD must be fine-tuned 
to $\bar{\theta} < 10^{-9}$ to avoid conflicting with the upper limit
on the neutron electric dipole moment.

In the decade of the 1980s, the areas of weak CP violation and strong CP
proceeded along largely separate tracks.

Having mentioned time-honored classics of the subject of CP, 
in the rest of the talk I shall specialize to six recent papers 
on a specific CP model - the aspon model - published: two\cite{aspon1,aspon2} 
in 1991, one\cite{aspon3}
in 1992, one\cite{aspon4} in 1994, one\cite{aspon5} in 1997, and
finally one in 1998\cite{aspon6}.

\section{Aspon Model}

Because QCD has a possible term involving $\bar{\theta}$ in its lagrangian,
there is the potential for unacceptably large CP violation. One approach
which is much less motivated now than 
twenty years ago is to introduce a color-anomalous U(1); a second is to
assume the up quark is massless, although this clashes with successes of
chiral perturbation theory. The third direction, exemplified by the aspon model
is to assume CP is a symmetry
of the fundamental theory and to arrange that $\bar{\theta}$ is zero
at tree level, remaining sufficiently small from radiative corrections.

In the aspon model the gauge group of the standard model is extended to
$SU(3) \times SU(2) \times U(1) \times U(1)_{new}$. The new charge
$Q_{new}$ is not carried by any of the fields of the SM. One additional
doublet of Dirac quarks (U, D) with charge $Q_{new}=1$ is introduced,
together with two complex singlet scalars $\chi^{\alpha}, \alpha=1,2$.

The $\chi^{\alpha}$ acquire VEVs with a non-zero relative phase,
spontaneously breaking both the gauged $U(1)_{new}$ and CP.
The gauge boson of $U(1)_{new}$ becomes massive by the Higgs mechanism
and is called the "aspon".

The Yukawa couplings with $\chi$ involve the right-handed U and D but
not the left-handed counterparts. As a result there are zeros\cite{NB} in the
$4 \times 4$ quark mass matrices such that although there are complex
entries the determinant is real. Hence $\bar{\theta} = 0$ at tree level. 

Such a mass matrix is diagonalized by a bi-unitary transformation which
is conveniently expanded in the small parameters $x_i = F_i/M$
where $F_i$ are the off-diagonal elements and M is the Dirac mass.
We may regard the $x_i$ as independent of the family number i and simply
write $|x_i| = x$. It turns out that x is constrained to lie in the window
$3 \times 10^{-5} < x^2 < 10^{-3}$ by the constraints of $\bar{\theta}$ and of 
CP violation. 

\subsection{FCNC}

Since we have introduced right-handed doublets, a first concern is with
the size of the induced Flavor-Changing Neutral Currents (FCNC). It
turns out that these are more than adequately suppressed.

\subsection{$\bar{\theta}$} 

At one loop level $\bar{\theta}$ acquires a non-zero value and this leads to
an upper limit on the product $(\lambda x^2)$ where $\lambda$ is the
coefficient of the quartic coupling $|\phi|^2|\chi|^2$ between the 
standard Higgs $\phi$ and the $\chi$ fields.

\subsection{Weak CP Violation}

Fitting to the CP violation parameter $\epsilon$ and to the allowed range
for $Re(\epsilon^{'}/\epsilon)$ gives an upper limit on the symmetry breaking
scale for $U(1)_{new}$ of about 2TeV. One thus predicts that, assuming the gauge
coupling is not much larger than the others of the standard model, the new
particles Q and A lie well below 1TeV. This fits ones intuition that if the
new states are too heavy the diagrams contributing to CP violation in the kaon system
will be too small.

\section{Production of A and Q at LHC}

Production of $\bar{Q}Q$ is dominated by gluon fusion diagrams just like $\bar{t}t$
production. The aspon A can be bremsstrahlunged from a heavy quark. Detailed calculations
show that the cross-section for aspon production is a few picobarns corresponding to
a few tens of thousands of events per year at LHC.

Of special interest is the decay width of A which depends sensitively on the A mass relative
to the Q mass M. For the most suppressed decay, when $M(A) < M(Q)$, the decay width
can be as small as 1KeV which is striking for a particle weighing several hundred GeV!

\section{B Decay}

The KM mechanism can be nicely checked from the unitarity triangle formed by the complex
numbers in the equation:
\begin{equation}   
V_{ub}^*V_{ud} + V_{tb}^*V_{td} + V_{cb}^*V_{cd} = 0
\end{equation} 
with corresponding angles $\alpha$, $\beta$, and $\gamma$. Using the expansion of the CKM matrix
as a power series in the Cabibbo angle\cite{wolfenstein}, it is profitable to define the
ratios:
\begin{equation}   
R_b = \left| \frac{V_{ud}^*V_{ub}}{V_{cd}^*V_{cb}} \right|
\end{equation} 
and 
\begin{equation}   
R_t = \left| \frac{V_{cd}^*V_{tb}}{V_{cd}^*V_{cb}} \right|
\end{equation} 
Clearly if the angle $\beta$, for example, is a significant value, well away from
zero or $\pi$( as would follow if the KM mechanism is the full explanation of the
CP violation in kaon decay), the $R_b+R_t > 1$.

It is well-known\cite{neubert} how to establish the angle $\beta$ from the expected data
on B decay coming from the B Factories under construction at SLAC and KEL Laboratories.

\subsection{CP Asymmetries in B Decay}

In the aspon model the $3 \times 3$ mixing matrix for the light quarks
is a real orthogonal one up to corrections of order $x^2$. This means that
the CP asymmetries of B decay are predicted to be at least three orders of
magnitude smaller than predicted by the KM mechanism.

In a general way, we may say that the KM mechanism is special in that the CP violation
in B decay is enhanced by a factor $(m_t/m_c)^2 \sim 10^4$ relative to that in K decay.
In most alternative models of CP violation such as the apon model, there is no reason
to expect this enhancement.

A clear prediction of the aspon model is that, to within less than $0.1\%$, $R_b+R_t=1$.
An unbiased study of the present data shows that this is well within the present
range.

\section{Kaon system reevaluated}

The value of $|\epsilon_K| = 2.26\times10^{-3}$ implies
(from aspon exchange) that
\begin{equation}
\kappa/x^2 = 2.8\times10^{3} GeV
\end{equation}
which, given the range for $x^2$, implies that $29TeV > \kappa > 870GeV$
from which the aspon mass is expected in the range 260GeV to 8.7TeV.

Contributions to $Re(\epsilon^{'}/\epsilon)$ come from tree diagrams
and penguin diagrams. A careful comparison to the
standard model gives a suppression of at least two orders
of magnitude. Consequently, observation of a value above $10^{-4}$
would exclude this model.

\section{Summary}

The main attractions of the aspon model are that it solves the strong CP problem,
accommodates weak CP violation, and makes testable predictions. A reader
who wishes to know more may consult the References listed.

\section*{Acknowledgments}
This work was supported in part by the US Department of Energy
under Grant No. DE-FG05-85ER-40219.

\end{document}